\documentstyle[12pt,epsf,psfig]{article} 
\begin{document} 
\title{\bf   Lepton flavor violation decays  with  the fourth
generation neutrino $\nu'$}
 \author{Wu-Jun Huo\,\,and \,\,Tai-Fu Feng\\{\sl CCAST (World Lab.),
P.O. Box 8730, Beijing $100080$}\\{\sl  and} \\ {\sl
Institute of High Energy Physics, Academia Sinica, P.O. Box $918(4)$},\\{\sl
 Beijing $100039$, P.R.  China}}

\date{today} 
\maketitle

\begin{abstract}
We investigate the lepton flavor violation decays, $\tau \to \mu\gamma$,
$\tau \to e\gamma$ and $\mu \to e\gamma$, in the
framwork of a squential fourth generation model with a heavy fourth neutrino,
$\nu'$. Using the recent experimental bounds, we can obtain the constraints 
of the  $4\times 4$ leptonic mixing  matrix element factors,
$V_{1\nu'},V_{2\nu'}$ and $V_{3\nu'}$. We find that LFV decays can give
strigent costraints on the parameter space of the 4th neutrino mass
$m_{\nu'}$.

\end{abstract}
\newpage

At present, Standard model (SM) has to face  the experimental difficaties
\cite{neutrino,g2} which are all relate to leptons. It seems to indicate the
presence of new physics just round the cornor will be in the leptonic part. 
The leptonic flavor violation decay $\tau\to \mu \gamma$ is very useful to
search new physics. 
In this note, we consider the sequential fourth generation
standard model (SM4) to investigate its contributions to  some leptonic flavor
violation decays, $\tau \to \mu\gamma$,
$\tau \to e\gamma$ and $\mu \to e\gamma$. 
 
   From the point of phenomenology,
 there is a realistic question what are numbers of the fermions
generation or weather there are other additional quarks or leptons. The
present experiments  tell us there are only three generation fermions with
$light$ neutrinos which mass are less smaller than $M_Z /2$\cite{Mark}. But
the experiments don't  exclude the existence of other additional generation,
such as the fourth generation, with a $heavy$ neutrino, i.e. $m_{\nu_4} \geq
M_Z /2$\cite{Berez}. Many refs. have studied  the SM4 \cite{McKay}, which is
added   an up-like quark $t^{'}$, a down-like quark $b^{'}$, a lepton
$\tau^{'}$,    and a heavy neutrino $\nu^{'}$ in the SM. The properties of
these new    fermions are all the same as their corresponding counterparts 
  of other three generations except their masses and CKM mixing, see Tab.1,
\begin{table}[htb] 
\begin{center} 
\begin{tabular}{|c||c|c|c|c|c|c|c|c|} 
\hline 
& up-like quark & down-like quark & charged lepton &neutral lepton \\ 
\hline 
\hline 
& $u$ & $d$& $e$ & $\nu_{e}$ \\ 
SM fermions& $c$&$s$&$\mu$&$\nu_{\mu}$ \\ 
& $t$&$b$&$\tau$&$\nu_{\tau}$\\
\hline
\hline new
fermions& $t^{'}$&$b^{'}$&$\tau^{'}$&$\nu'$ \\ 
\hline 
\end{tabular}
\end{center}
\caption{The elementary particle spectrum of SM4} 
\end{table}

If there exists a very heavy fourth neurino $\nu'$,  it can contribute to
$a_\mu$ through diagram of Fig. 1.
This is the electroweak interaction. Similar to that of quarks, the
corresponding Lagragian is 
\begin{equation}
{\cal L}=-\frac{g}{\sqrt{2}}({\bar \nu'}\gamma_\mu a_L V_{l\nu'} l) W^\mu
+h.c.
\end{equation}
where $a_L =(1-\gamma_5)/2$, $V_{l\nu'}$ ($l=e,\mu,\tau$ )is the (1,4), (2,4)
and (3,4) elements of the four-generaton CKM matrix ($4\times 4$),
 \begin{equation}
V^{\rm SM4}_{\rm CKM} = \left (
\begin{array}{lcrr}
V_{1\nu_e} & V_{1\nu_\mu} & V_{1\nu_\tau} & V_{1\nu'}\\
V_{2\nu_e} & V_{2\nu_\mu} & V_{2\nu_\tau} & { V_{2\nu'}}\\
V_{3\nu_e} & V_{3\nu_\mu} & V_{3\nu_\tau} & V_{3\nu'}\\
V_{4\nu_e} & V_{4\nu_\mu} & V_{4\nu_\tau} & V_{4\nu'}\\
\end{array} \right )
\end{equation}   

Reverting back to the diagrams of Fig. 1, we see that the fourth neutrino contribution to
decays $\tau \to e\gamma, \mu\gamma$ and $\mu\to e\gamma$ (see Fig. 1). We
don't consider the usual three neutrinos' contribution because their masses
are too small. For the heavy  neutrino $\nu'$, In SM4, it can induce thses
decaies. Calculating Fig. 1, one obtains 
\begin{equation}
{\rm \Gamma}(\mu\to e\gamma) =\frac{\alpha G_F^2 m^5_\tau}{96}
|V_{1\nu'} V_{2\nu'}|^2 f^2  (x),
\label{taumu}
\end{equation}
where $x\equiv m^2_{\nu'} /m^2_W$, $\alpha$ is the fine constrcture constant
and 
\begin{eqnarray}
 f(x)=\frac{-5x^3 -5x^2 +4x}{12(x-1)^3}+\frac{(2x^3 -x^2 )\log x}{2(x-1)^4}.
\end{eqnarray}
Simlarly, we obtain
\begin{equation}
{\rm \Gamma}(\tau\to e\gamma) =\frac{\alpha G_F^2 m^5_\tau}{96}
|V_{1\nu'} V_{3\nu'}|^2 f^2  (x),
\label{taue}
\end{equation}
\begin{equation}
{\rm \Gamma}(\tau\to \mu\gamma) =\frac{\alpha G_F^2 m^5_\mu}{96}
|V_{2\nu'} V_{3\nu'}|^2 f^2  (x).
\label{mu}
\end{equation}
Using the current experimental bounds of these three LFV processes \cite{pdg},
\begin{eqnarray}
{\rm Br}(\mu\to e\gamma) \leq 4.9 \times 10^{-11},\nonumber\\
{\rm Br}(\tau\to e\gamma) \leq 2.7 \times 10^{-6},\nonumber\\
{\rm Br}(\mu\to e\gamma) \leq 3.0 \times 10^{-6}.
\end{eqnarray}
qwe can obtain the parameter space of other $4\times 4$  matrix elements
factors of leptonic mixing, $|V_{l\nu'} V_{l'\nu'}|^2$ (see Figs. 2, 3, 4).

From the Figs, considering the unitarity of the matrix 
$V^{SM4}_{CKM}$, we find the reasonable  range of $m_\mu'$ is under teh
curve. But in our previous paper \cite{huo1}, the upper bounds of
$m_{\nu'}$ don't exceed  80$GeV$ from the revised $g_\nu -2$ results.  Although
the low bound of  $m_{\mu'}$ we get from $l\to l'\gamma $ is consistent with
the present  experiments \cite{Mark}, the upper bound of $m_{\mu'}$
 seems to conflict with the current experiments statue which  there is
no any new physics signals upper to hadrds GeVs. The  fourth generation
particles seems not to be so light. They should be several hadrad GeVs weight.

   In summary, we calculate the contribution of the fourth generation 
to LFV decays and get an interesting result which we can exclude most 
values of $m_{\nu'}$.  Considering the current experimental bounds,
 we give the parameter space of $m_{\mu'}$ and lepton mixing 
matrix element $V_{l\mu'}$. We find that LFV processes can constrain on 
the neutrino mass of the fourth generation: i.e. its mass should 
be lighter than 80 GeV. It seems that from 
the lepton part, the current experiments can impose a stringent 
constraint on the existence of the fourth generation.

\section*{Acknowledgments}
This research is supported  by the the Chinese Postdoctoral 
Science Foundation and CAS K.C. Wong Postdoctoral Research Award  Fund.
 We are grateful to Prof. X.M
Zhang for useful discussions.

\newpage
\begin{figure}
\vskip 14cm
\epsfxsize=20cm
\epsfysize=10cm
\centerline{
\epsffile{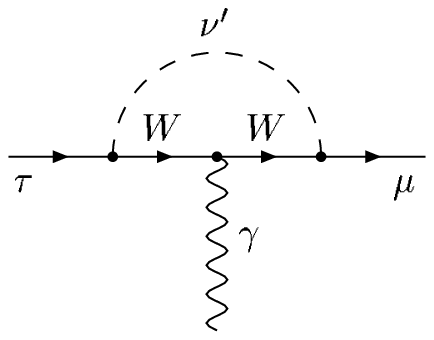}}
\vskip -18cm
\caption{ Feynmann diagram for LFV decay $\mu \to e\gamma$ induced by $\nu'$.}
\end{figure}

\newpage
\begin{figure}
\vskip 7cm
\epsfxsize=20cm
\epsfysize=10cm
\centerline{
\epsffile{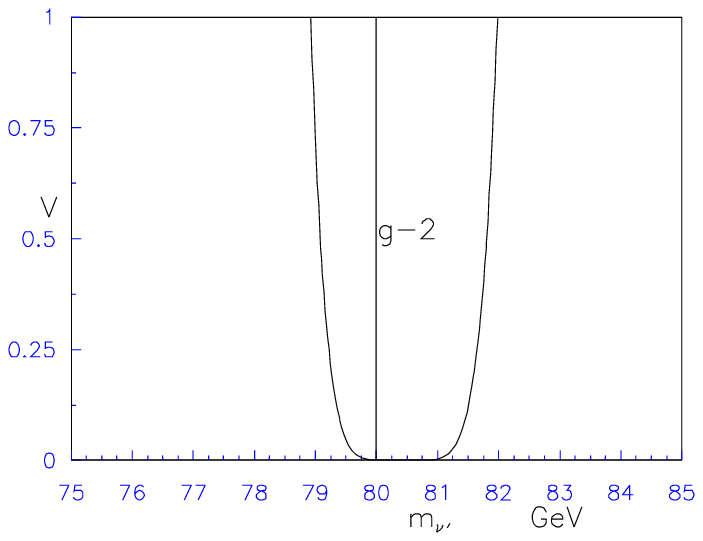}}
\caption{ Diagram of  $|V_{1\nu'} V_{2\nu'}|^2$ to $m_{\nu'}$ }
\end{figure}

\newpage
\begin{figure}
\vskip 7cm
\epsfxsize=20cm
\epsfysize=10cm
\centerline{
\epsffile{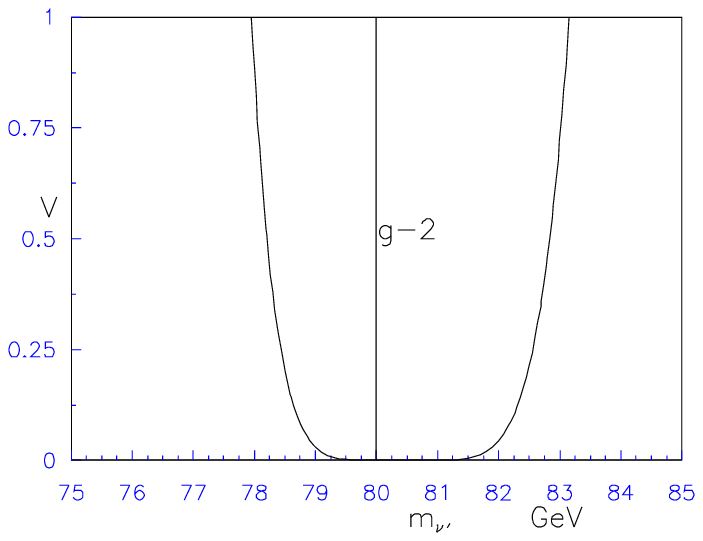}}
\caption{ Diagram of $|V_{1\nu'} V_{3\nu'}|^2 $ to $m_{\nu'}$ .}
\end{figure}

\newpage
\begin{figure}
\vskip 7cm
\epsfxsize=20cm
\epsfysize=10cm
\centerline{
\epsffile{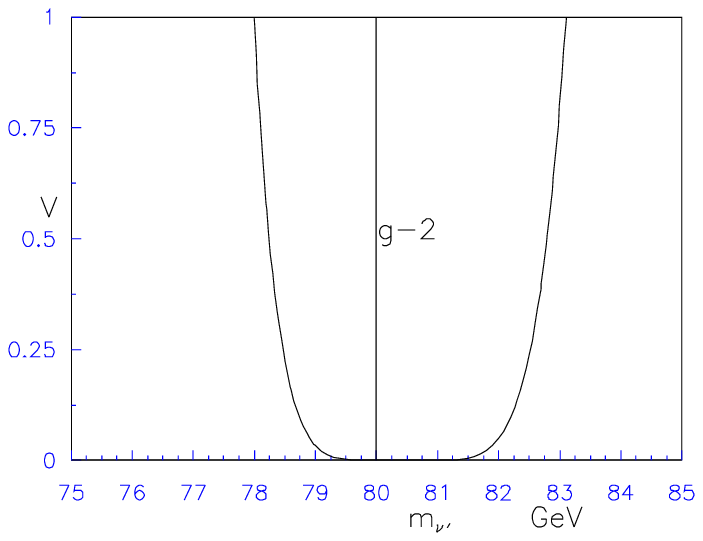}}
\caption{ Diagram of $|V_{3\nu'} V_{2\nu'}|^2$  to $m_{\nu'}$ .} \
\end{figure}

\end{document}